\documentclass{article}

\usepackage{PRIMEarxiv}

\usepackage[utf8]{inputenc}
\usepackage[T1]{fontenc}

\usepackage{amsmath,amssymb,amsfonts}
\usepackage{gensymb}
\usepackage{dsfont}
\usepackage{nicefrac}
\usepackage{siunitx}

\usepackage{algorithmic}

\usepackage{graphicx}
\usepackage[inkscapelatex=false]{svg}
\graphicspath{{media/}}

\usepackage{booktabs}
\usepackage{tabularx}
\usepackage{array}

\usepackage{textcomp}
\usepackage{microtype}
\usepackage{lipsum}

\usepackage{fancyhdr}

\usepackage{cite}
\usepackage{url}
\usepackage{hyperref}

\usepackage[capitalise]{cleveref}

\DeclareMathOperator*{\argmin}{arg\,min}

\newcommand{\ul}[1]{\underline{#1}}
\newcommand{\matr}[1]{\mathbf{#1}}
\newcommand{\ubeta}{\ul{\beta}}
\newcommand{\vv}{\ul{v}}
\newcommand{\vY}{\ul{Y}}
\newcommand{\vy}{\ul{y}}
\newcommand{\vX}{\ul{X}}
\newcommand{\vV}{\ul{V}}
\newcommand{\vx}{\ul{x}}
\newcommand{\vh}{\ul{h}}
\newcommand{\vH}{\ul{H}}
\newcommand{\vu}{\ul{u}}
\newcommand{\vw}{\ul{w}}
\newcommand{\vz}{\ul{z}}
\newcommand{\vmu}{\ul{\mu}}
\newcommand{\vlambda}{\ul{\lambda}}
\newcommand{\aux}[1]{\widetilde{#1}}
\newcommand{\auxd}{\tilde{d}}

\newcommand{\mw}{\matr{w}}

\newcommand{\hmw}{\widehat{\matr{w}}}

\newcommand{\mz}{\matr{z}}

\newcommand{\hmz}{\widehat{\matr{z}}}
\newcommand{\mx}{\matr{x}}
\newcommand{\mX}{\matr{X}}
\newcommand{\hmx}{\widehat{\matr{x}}}

\newcommand{\mG}{\matr{G}}
\newcommand{\mlambda}{\matr{\lambda}}
\newcommand{\mgamma}{\matr{\gamma}}

\newcommand{\hlambda}{\hat{\lambda}}
\newcommand{\htheta}{\hat{\theta}}
\newcommand{\htau}{\hat{\tau}}
\newcommand{\hZ}{\hat{Z}}

\newcommand{\reals}{\mathbb{R}}

\newcommand{\E}{\mathbb{E}}

\newcommand{\SPCM}[3]{\mathcal{C}_{{#1}\rightarrow {#2}}[#3]}

\title{Efficient Convex Optimization Methods for State-Space Heartbeat Dynamics Models with Gamma Generalized Linear Models
}

\author{
  Sabrina Liu \\
  Department of Electrical Engineering \\
  Stanford University \\
  Stanford, CA, U.S.A.\\
  \texttt{sabliu@stanford.edu} \\
   \And
  Andrew S. Perley \\
  Department of Neurology \\
  Massachusetts General Hospital \\
  Boston, MA, U.S.A.\\
  \texttt{aperley@mgh.harvard.edu} \\
  \AND
  Todd P. Coleman \\
  Department of Bioengineering \\
  Stanford University \\
  Stanford, CA, U.S.A. \\
  \texttt{toddcol@stanford.edu} \\
}

\begin{document}
\maketitle

\begin{abstract}
Objective: This work proposes a dynamic state-space statistical model of heartbeat dynamics whose optimal state estimate can be solved efficiently with convex optimization and used to assess autonomic and cardiovascular health. Methods: This state-space model combines a Gauss-Markov process prior with gamma emissions using a log-link function. Parameters are estimated using a static version of the model, and alternating direction method of multipliers is used to efficiently find the optimal maximum a priori latent state estimate. Model selection is performed by finding the tradeoff between model size and model fit using the partial autocorrelation function. Results: We demonstrate in synthetically generated heartbeats that this model is able to closely recover the underlying latent weights. Then, we show using a tilt table dataset that this statistical model is able to better capture sharp transients in heart rate patterns than traditional local averaging methods. In addition, combining heart beat timings with information from other organs such as the respiratory phase, we show that our statistical modeling framework enables the extraction of information-theoretic measures to quantify inter-organ coupling such as respiratory sinus arrythmia. Conclusion: This statistical framework for modeling heartbeat dynamics provides an efficient framework that can provide physiologically meaningful insights. Significance: This work broadens potential tools that can be used to assess heartbeat dynamics.
\end{abstract}

% keywords can be removed

\section{Introduction}
\label{sec:introduction}
Statistical modeling of the timings of heartbeats can provide insight into the autonomic processes that control them. This can provide value during in-clinic visits for diagnosing a large spectrum of autonomic conditions, such as orthostatic hypotension \cite{lanier2011evaluation}. As heart rate data becomes increasingly available with the advent of wearables, ambulatory applications are also of immediate interest, such as in stress management \cite{mezzacappa2001vagal} and exercise \cite{cole1999heart}.

There are several preexisting statistical models for such features. Several consider static dynamics and assume that the latent processes that describe these dynamics are time-invariant. While this is a reasonable assumption for short-term recordings at rest, when the heart is responding to various stimuli or during extended recordings, such assumptions no longer apply. Prior work has incorporated time-varying dynamics to encode how the relationship between heartbeats and their history can change over time \cite{barbieri2005analysis}. Model-fitting procedures for these models, however, are often nonconvex and slow to solve, resulting in difficulty in adoption. This work will overcome these challenges by using a state-space, point-process model whose optimal estimate can be obtained using an efficient convex optimization procedure. Compared to previous models, while we lose some physiological interpretability, this formulation provides computational efficiency.

We will begin in \cref{sec:relatedwork} by discussing previous related work in the field of modeling heartbeat dynamics. Then, in \cref{sec:methods} we will go into detail about our proposed model and describe an efficient procedure for estimating the model's parameters. In \cref{sec:results}, we demonstrate its performance on synthetic heartbeats to show its ability to accurately estimate the ground truth latent states. We also analyze data from a tilt table experiment and demonstrate our model's abilities to capture sharp transient changes in heartbeat dynamics. Finally, using simultaneous cardiac-respiratory data, we show how the effects of additional regressors like respiration can be used to quantify physiological phenomena such as respiratory sinus arrythmia. We provide a discussion and consider next steps in \cref{sec:discussion}. Altogether, we hope that this model can provide an additional resource for clinicians and engineers to statistically model and quantify heartbeat dynamics. A preliminary version of this work has been reported in \cite{liu2024efficient}.

\begin{figure}[t]
    \centering
    \includegraphics[width=\columnwidth]{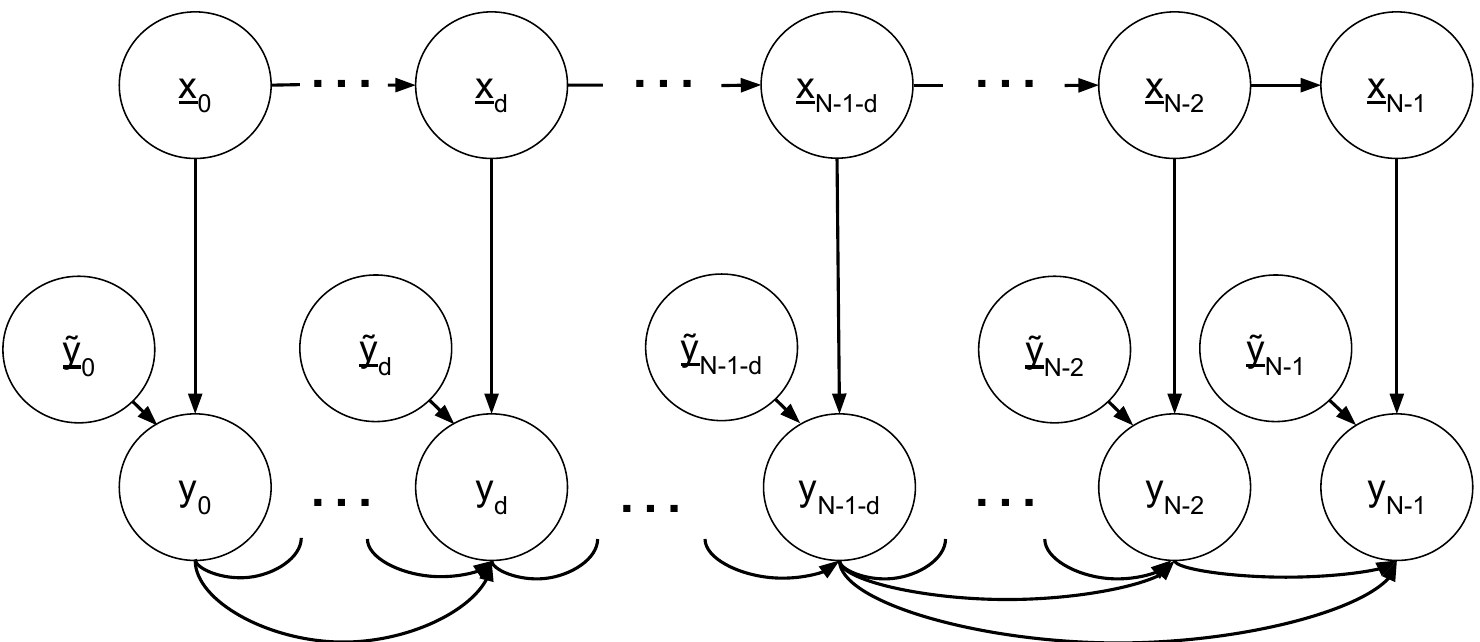}
    % \includesvg[width=\columnwidth]{figures/journal_sshbd.svg}
    \caption{State-space model for heartbeat dynamics. $y_n$ is the observed $n$th RR interval. $\vx_n$ is the $n$th latent vector of weights that determines the affine combination relating the past $d$ intervals to the current one.}
    \label{fig:sshbd}
\end{figure}

\section{Related work}
\label{sec:relatedwork}
Traditionally, measures of heartbeat dynamics, such as mean heart rate and heart rate variability (HRV), rely on simple statistical measures on windowed data \cite{shaffer2017overview}. Spectral methods have also been used to break down and associate different frequency bands with autonomic regulation of the heart \cite{malliani1991cardiovascular}. Alternative representations to better distinguish parasympathetic and sympathetic dynamics have been developed using other bases functions \cite{valenza2018measures}. Nonlinear approaches based on fractal dynamics \cite{ivanov1999multifractality} and coupled oscillators \cite{dos2004rhythm} have also been shown to be useful in characterizing rhythmic differences between health and pathology.

Point-process models are another set of methods that instead treats heartbeats as discrete events and directly models the timings between them. They have been used in other physiological contexts to describe the spiking of neurons \cite{truccolo2005point} and electrodermal activity \cite{subramanian2020point}. In the cardiac setting, \cite{barbieri_pointprocess_2005} models the timings between heartbeats, also known as RR intervals, as a history-dependent inverse Gaussian (IG) distribution to capture the dependencies between heartbeats. The model assumes this autoregressive nature remain stationary over time. A state-space model (SSM) was later used to provide flexibility in the model to capture autonomic changes \cite{barbieri2005analysis}. While the parameters of this model can be solved with an adaptive filtering procedure, the non-convexity in the model-fitting procedure can make it slow to fit and result in suboptimal solutions.

Prior work has since adapted the static point-process model to use a gamma distribution with a log-link function \cite{perley2025convex}. While losing some physiological interpretation and having worse model fit compared to the IG distribution \cite{chen2009assessment}, the set of distributions modeled by these changes still have many of the same physiological characteristics seen in RR interval distributions, namely that they are unimodal, have a long right tail, and have positive support. In addition, this alternate modeling framework with gamma distributions also affords convexity in the model-fitting procedure to optimally estimate model parameters.

In this work, we expand this latter convex model by incorporating an SSM to reflect changes in the autonomic nervous system (ANS). To do this, we impose a Gauss-Markov prior (GMP) on our model parameters and use a generalized linear model (GLM) to parameterize the gamma distributions of the emissions. With this new model, we describe a convex optimization procedure called alternating direction method of multipliers (ADMM) to optimally estimate its latent weights. In addition, we propose a method using partial autocorrelation to determine the model order. Goodness-of-fit is assessed using the time-rescaling theorem. In synthetic data as well as clinical recordings, we demonstrate how the model can capture changes in heartbeat dynamics under different stimuli. In addition, as an expansion from previous such models, we show how we can incorporate additional regressors into our model and use information-theoretic measures to quantify inter-organ effects such as respiratory sinus arrhythmia.
\section{Methods}
\label{sec:methods}
\begin{figure*}[t]
    \centering
    \includegraphics[width=\textwidth]{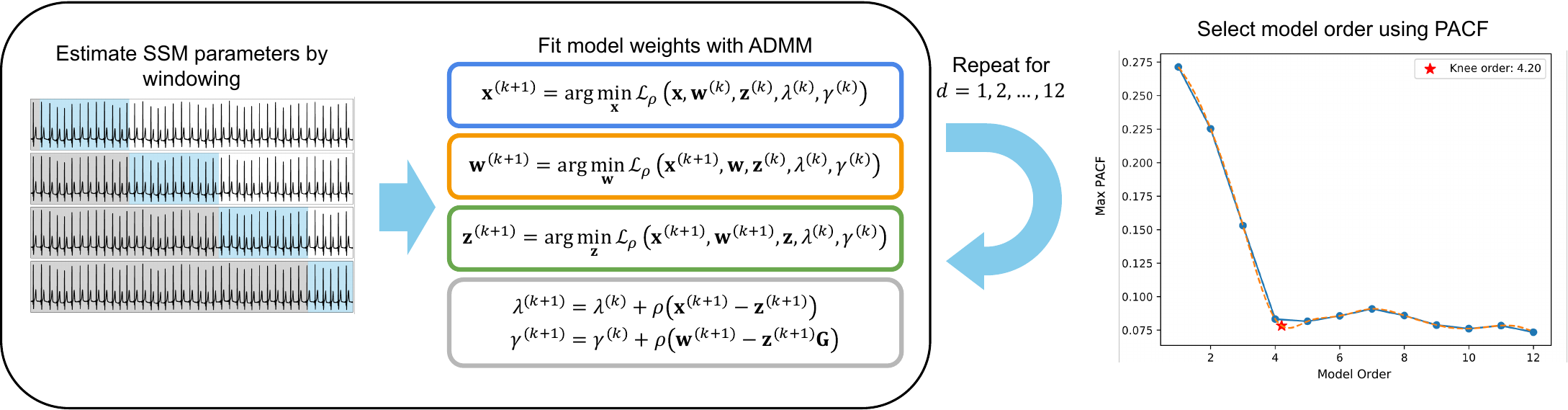}
    % \includesvg[width=\textwidth]{figures/procedure_conceptual.svg}
    \caption{Proposed model fitting procedure. Model parameters are estimated by fitting the static model to windows of the data. From there, ADMM is used to solve for the latent weights. After repeating this procedure for all possible model orders, trends in the PACF are used to identify the desired model order.}
    \label{fig:modelFitProcedure}
\end{figure*}
We begin by describing our proposed model, conceptually shown in \cref{fig:sshbd}, as well as a procedure for parameter estimation and model selection. The steps for model fitting are illustrated in \cref{fig:modelFitProcedure}.

\subsection{Notation}
In this section, we denote a column vector by underlining the variable (e.g., $\vv$) and a matrix by bolding the variable (e.g., $\matr{x}$). We denote a random variable $X$ with capitalization and a specific realization $x$ with lower case. $<\cdot, \cdot>_F$ and $||\cdot||_F$ denote the Frobenius inner product and the Frobenius norm, respectively.

\subsection{Proposed model}
The proposed model aims to capture the autoregressive nature of heartbeat intervals. In addition, it can also capture the influence from other systems such as other organs or external stimuli. 

To model this, consider $N$ interbeat intervals as the random vector $\vY = [Y_0,\ldots,Y_{N-1}]^T$.  Other stimuli that may influence $\vY$ are represented by $\aux{\vY} = [\aux{\vY}_0,\ldots, \aux{\vY}_{N-1}]^T \in \reals^{\auxd \times N}$. We allow $\aux{\vY}_n \in \reals^{\auxd}$ to be multivariate to capture multiple regressors at each time step.

We assume there is a latent Gauss-Markov process (GMP) $\matr{X} = [\ul{X}_0,\ldots, \ul{X}_{N-1}]$ given by 
\begin{eqnarray}
\vX_{n+1} = \vX_n + \vV_n \qquad \forall n \geq 1, \label{eqn:defn:stateprocess}
\end{eqnarray}
where $d$ is the autoregressive model order and $\auxd$ is the dimension of auxiliary regressors. Therefore, $\ul{X}_n$ is of dimension $d+\auxd+1$. The distribution of the initial state is described by $\vX_0 \sim \mathcal{N}(\vmu_0, \Sigma_0)$, and the transition matrix between states is described by $\vV_n \sim \mathcal{N}(\ul{0},\Sigma_v)$. This captures time-varying changes in the underlying distribution of interbeat intervals. We describe $f_{\vX_n|\vX_{n-1}}(\vx_n|\vx_{n-1})$ succinctly as $p_n(\vx_n|\vx_{n-1})$. Note that for any $\vu$, $-\log p_n(\vv|\vu)$ is a quadratic form with a symmetric positive definite matrix and thus convex in $\vv$.

We consider a measurement model that captures the history-dependent properties of heartbeats, i.e. for each RR interval, its statistics are dependent on the recent RR intervals before it and other non-cardiac stimuli. We use gamma distributions which are similar to RR interval distributions in that they have positive support, are unimodal, and are right-tailed.

Formally, the $d$-th order model
$f_{Y_n|\vX_n,\vY_{n-1},\aux{\vY}_{0:n-1}}(\cdot|\vx_n,\vy_{0:n-1},\aux{\vy}_{n-1})$ is only a function of the current state $\vx_n$, the previous $d$ RR intervals, and the observed additional regressors:
\begin{equation*}
    f_{Y_n|\vX_n,\vY_{n-1}, \aux{\vY}_{n}}(y_n|\vx_n,\vy_{0:n-1},\aux{\vy}_{n}) \equiv q_n(y_n|\vx_n,\vh_n),
\end{equation*}
where $\vh_n = [1, y_{n-1}, \dots, y_{n-d}, \aux{y}_{0}, \dots \aux{y}_{\auxd-1}]^T$.
Therefore, we let the above conditional distribution be a gamma distribution with parameters $\alpha$ and $\beta_n$, which we specify with a GLM to encode the history dependence:
\begin{align}
    \log\frac{\alpha}{\beta_n} &= \log \E \left[Y_n|\vX_n = \vx_n, \vH_n = \vh_n \right]\nonumber \\
    &= \ul{h}_n^T \vx_n \label{eqn:meanRR}.
\end{align}
Note that for any fixed $\alpha$, $\beta_n$ depends on the previous $d$ RR intervals through a log-link function of the mean, given by $\alpha/\beta_n$ for gamma distributions.
$-\log q_n(y_n|\vx_n,\vh_n)$ is convex in $\vx_n$, given by
\begin{align}
-\log q_n(y_n|\vx_n, \vh_n) &= \log \Gamma(\alpha) -\alpha \log \alpha - (\alpha-1) \log y_n \nonumber \\
&+\alpha y_n e^{- \vh_n^T \vx_n} + \alpha  \vh_n^T \vx_n.
\label{eqn:defn:NLL}
\end{align}
This fact, along with how $-\log p_n(\vx_n|\vu)$ is convex in $\vx_n$, implies log-concavity of the posterior distribution $f_{\matr{X}|\vY}(\cdot|\vy)$. Another attractive property of the gamma GLM in \eqref{eqn:defn:NLL} with the latent GMP in \eqref{eqn:defn:stateprocess} is that it is self-consistent. Specifically, each RR interval $Y_n$ is nonnegative, and we link its conditional expectation to an affine transformation of the latent GMP (which can be positive or negative) via the log-link function in \eqref{eqn:meanRR}. 
% In contrast, the inverse Gaussian model in \cite{barbieri2005analysis} equates the conditional expectation of the positive RR intervals to an affine transformation of the (possibly negative) GMP.

Our goal is to solve for the time-varying weights \\$\matr{X} \in \reals^{(d+\auxd+1)\times N}$
 and the parameters $\alpha$, $\vmu_0$, $\Sigma_0$, and $\Sigma_v$.

\subsection{Solving for the time-varying, latent weights}
Assuming all other parameters are known, the MAP estimate of the weights is given by:
\begin{align}
        \hmx_{\text{MAP}} =& \argmin_{\mx \in \reals^{(d+\aux{d}+1) \times N}}  - \log f_{\mX|\vY}(\mx|\vy) \nonumber \\
         =&  \argmin_{\mx \in \reals^{(d + \aux{d}+1) \times N}}-\log p_0(\vx_0) -\sum_{n=1}^{N-1} \log p_{n}(\vx_n | \vx_{n-1}) \nonumber\\ 
        & \qquad - \sum_{n=0}^{N-1} \log q_n(y_n|\vx_n, \vh_n).
    \label{eqn:ogOptProblem}
\end{align}

Even though the model-fitting procedure to find $\hmx_{\text{MAP}}$ in \eqref{eqn:ogOptProblem} is a convex optimization procedure, using black box convex solvers such as CVXPY \cite{diamond2016cvxpy} can be infeasible when the number of samples $N$ is large. 
Instead, we use the convex optimization procedure alternating direction method of multipliers (ADMM) \cite{boyd2011distributed}. Based on the framework in \cite{schamberg2018modularized}, we introduce auxiliary variables $\mw$ and $\mz$ to separate terms in the  objective function:
\begin{align*}
    (\hmx, \hmw, \hmz) &= \argmin_{\mx,\mw,\mz} -\log p_0(\vw_0) - \sum_{n=1}^{N-1} \log p_n(\vw_n)\\ 
    & \qquad - \sum_{n=0}^{N-1} \log q_n(y_n|\vx_n, \vh_n)\\
    \text{s.t. } \vx_n &= \vz_n \qquad \qquad \quad \; \forall n \in \{0, \dots, N-1\}\\
    \vw_0 &= \vz_0 \\
    \vw_n &= \vz_n - \vz_{n-1} \qquad \forall n \in \{1, \dots, N-1\}.
\end{align*}
We form the augmented Lagrangian, with a slight abuse of notation for the priors:
\begin{equation}
    \begin{split}
    \mathcal{L}_{\rho}&(\mx, \mw, \mz, \mlambda, \mgamma) = \\
    &-\log p_0(\vw_0) - \sum_{n=1}^{N-1} \log p_n(\vw_n) \\
    &- \sum_{n=0}^{N-1} \log q_n(y_n|\vx_n, \vh_n)\\
    &+ \left<\matr{\lambda}, \mx-\mz\right>_{F} + \left<\mgamma, \mw - \mz\mG\right>_{F}\\
    &+ \frac{\rho}{2} ||\mx-\mz||_F^2 + \frac{\rho}{2}||\mw - \mz\mG||_F^2
     \end{split}
    \label{eqn:lagrangian}
\end{equation}
where $\mG \in \reals^{N \times N}$ is a finite difference matrix:
\begin{equation*}
    \mG = \begin{bmatrix}
        1 & -1 & & 0 \\
         & \ddots & \ddots & \\
         & & 1 & -1 \\
        0 &  & & 1
    \end{bmatrix}.
\end{equation*}
We can then optimize \eqref{eqn:lagrangian} via dual coordinate ascent:
\begin{subequations}\label{eqn:ADMM}
\begin{align}
    &\mx^{(k+1)} = \argmin_{\mx} \mathcal{L}_{\rho}(\mx, \mw^{(k)}, \mz^{(k)}, \mlambda^{(k)}, \mgamma^{(k)}) \label{eqn:xUpdates}\\
    &\mw^{(k+1)} = \argmin_{\mw} \mathcal{L}_{\rho}(\mx^{(k+1)}, \mw, \mz^{(k)}, \mlambda^{(k)}, \mgamma^{(k)}) \label{eqn:wUpdates}\\
    &\mz^{(k+1)} = \argmin_{\mz} \mathcal{L}_{\rho}(\mx^{(k+1)}, 
    \mw^{(k+1)}, \mz, \mlambda^{(k)}, \mgamma^{(k)}) \label{eqn:zUpdates}\\
    &\mlambda^{(k+1)} = \mlambda^{(k)} + \rho (\mx^{(k+1)} - \mz^{(k+1)}) \label{eqn:lamUpdates}\\
    &\mgamma^{(k+1)} = \mgamma^{(k)} + \rho (\mw^{(k+1)} - \mz^{(k+1)}\mG)\label{eqn:gammUpdates}
\end{align}
\end{subequations}

Of these updates, except for \eqref{eqn:xUpdates}, all are linear and can be solved with matrix operations.

As for \eqref{eqn:xUpdates}, it is parallelizable and each coordinate $\vx_n$ can be solved for separately. Instead of using a generic convex solver, computations can be further sped up by using the solution to the Lambert function. To do this, we note each coordinate in \cref{eqn:xUpdates} is given by:
\begin{equation}
    \vx_n^{(k+1)} = \argmin_{\vx \in \reals^{d+1}}\;\alpha y_n e^{-\vh_n^T \vx} + \alpha \vh_n^T \vx + \frac{\rho}{2}||\vx||_2^2 - (\rho \vz - \vlambda)^T \vx
\end{equation}

Since we know this expression is convex, we can find the minimum by taking the gradient of the objective with respect to $\vx$ and setting it to $0$:
\begin{equation}
\begin{split} 
    \nabla_{\vx}(\alpha y_n &e^{-\vh_n^T \vx} + \alpha \vh_n^T \vx + \frac{\rho}{2}||\vx||_2^2 - (\rho \vz - \vlambda)^T \vx) \\
    &= -\alpha y_n e^{-\vh_n^T\vx} \vh_n + \alpha \vh_n + \rho \vx - \rho \vz + \lambda = 0
\end{split}
\end{equation}

Reordering the terms and multiplying them all by $\vx_n$, we get:
\begin{equation}
    \label{eqn:lambertProblem}
    \begin{split}
    \vh_n^Tx &= \vh_n^T \vz - \frac{1}{\rho}\vlambda^T \vh_n - \frac{\alpha}{\rho} \vh_n^T \vh_n + \frac{\alpha y_n}{\rho}(\vh_n^T \vh_n) e^{-\vh_n^T\vx} \\
    w &= a + be^{cw}
    \end{split}
\end{equation}
where we let 
\begin{align*}
    a &= \vh_n^T z - \frac{1}{\rho} vh_n^T \vlambda_n - \frac{\alpha}{\rho}\vh_n^T \vh_n \\
    b &= \frac{\alpha y_n}{\rho}\vh_n^T \vh_n \\
    c &= -1\\
    w &= \vh_n^T \vx
\end{align*}

\Cref{eqn:lambertProblem} has no closed form solution for $w$. However, it can be seen as the intersection of a line and an exponential function in $w$. The solution for these intersections are the solutions to the Lambert function $W$ \cite{corless1996lambert}. We can thus write a solution for $w$ as:
\begin{equation}
    \label{eqn:lambertAnswer}
    w = a - \frac{1}{c}\mathcal{W}(-bce^{ac})
\end{equation}

where $\mathcal{W}$ is the Lambert function. Combining this with \cref{eqn:lambertProblem} gives us the solution to each coordinate as:
\begin{equation*}
    \vx_n^{(k+1)} = \frac{\alpha}{\rho} y_n e^{-w}\vh_n - \frac{\alpha}{\rho}\vh_n + \vz_n - \frac{1}{\rho}\vlambda_n.
\end{equation*}

For the analyses in this work, ADMM was run with $\rho=100$ until it reached convergence determined by satisfying the Cauchy criterion for all primal and dual variables or within $1000$ iterations. 

The coordinate ascent updates formulated according to \eqref{eqn:xUpdates} - \eqref{eqn:gammUpdates} also have a statistical interpretation: $\mx$ considers constraints imposed by the  gamma GLM measurement model, $\mw$ considers constraints imposed by the latent GMP prior, and $\mz$ seeks consensus between $\mx$ and $\mw$. Because of the log-concavity of the likelihoods and priors, we are guaranteed that ADMM will result in the optimal MAP estimate \cite[Thm. 1]{schamberg2018modularized}.

\subsection{Solving for the model parameters}
The previous framework to estimate the latent weights assumes that the model parameters $\theta = (\alpha,\vmu_0,\Sigma_0,\Sigma_v)$ are known. For the shape parameter of the gamma distribution, we fix $\alpha$ so that temporal dynamics and inter-subject variability are instead reflected primarily in the latent weights. Therefore, we treat it as a nuisance parameter and use a method-of-moments estimate based on reported values for typical human mean heartrate and HRV \cite{shaffer2017overview}. This gives us:
% To empirically estimate the shape parameter $\alpha$, we use a method of moments estimator based on the equations for the mean and variance of a gamma distribution in terms of its shape and rate parameters:
\begin{equation}
    \hat{\alpha} = \frac{\mu_{RR}^2}{\sigma_{RR}^2} = \frac{(\SI{1}{\second})^2}{(\SI{50}{\milli\second})^2} = 400
\end{equation}
where $\mu_{RR}$ and $\sigma_{RR}$ are the typical mean and standard deviation $RR$ interval length.

The other parameters are estimated using the maximum likelihood estimates (MLE) from the static heartbeat dynamics model in \cite{perley2025convex}. $\vmu_0$ is set as $\hat{\vx}_{M\!L\!E}$. $\Sigma_0$ and $\Sigma_v$ are computed by taking $100$ sliding window of $100$ intervals and computing $\hat{\vx}_{M\!L\!E}$ of each. $\Sigma_0$ is the empirical covariance of these estimates, while $\Sigma_v$ is the empirical covariance of the differences of these estimates.

\subsection{Solving for the model order}
A remaining question in our model is finding the optimal $d$ to capture the history-dependent nature of heartbeat dynamics. If $d$ is too small, then not enough prior heartbeats are selected to inform the inference of the current heartbeat distribution. On the other hand, if $d$ is too large, then the model may overfit to the data and not generalize well.

Different information criterion such as Akaike Information Criterion (AIC) \cite{akaike2025akaike} and Bayesian Information Criterion (BIC) \cite{schwarz1978estimating} are often used to find optimal model orders. However, in our model, the number of unknowns grows with the number of samples in our data. In such overparameterized models, AIC and BIC are known to be insufficient for model order selection \cite{hurvich1989regression, chen2008extended}.

As an alternative, another way to assess model fit and choose the order $d$ is to use the partial autocorrelation function (PACF) of the residuals. This has been used to assess model fit for generalized autoregressive moving average (GARMA) models and involves checking if the magnitude of the residuals is relatively small and if there are any prominent trends \cite{benjamin2003generalized}.

To implement this, we assume $d$ is between $1$ and $12$ based on prior autoregressive heartbeat dynamics models \cite{barbieri_pointprocess_2005}. For each value of $d$, we fit the model according to the procedure described previously in this section. We then compute the PACF based on the residuals for lags $1$ to $60$ (which is a maximum lag of around 1 minute). We find the maximum absolute PACF of all the lags and record that for each model order $d$. Autocorrelation in the residuals should be small if the model order $d$ adequately captures the temporal dependence between RR intervals. However, an overly large model order $d$ would unnecessarily increase model complexity and potentially overfit to data. Therefore, to systematically find the tradeoff point where this occurs, we use the Kneedle algorithm \cite{satopaa2011finding}. This procedure interpolates the data and then finds the point of maximum curvature. We then round the value of $d$ that achieved this point to the closest integer and use that as the optimal model order.

\subsection{Validation}
We assess model goodness of fit using the time-rescaling theorem, which was originally used for goodness-of-fit analyses of neural spike trains \cite{brown2002time}. Let $0 < u_0 < u_1 < \dots < u_{N-1}$ be the heartbeat timings.
We define $q_n(y_n|\vx_n;\theta)$ as defined in \eqref{eqn:defn:NLL} where $\theta$ indicates the model parameters. For $t \in [u_n, u_{n+1})$, the conditional intensity $\lambda_t$ of the heartbeat point process is given by \cite{brown2002time}:
\begin{equation}
    \lambda_t = q_n(t|\vX_n, \vh_n; \theta)\left[1 - \int_{u_n}^t q_n(s|\vX_n, \vh_n;\theta) ds\right]^{-1} \nonumber%\label{eqn:defn:lambdat}
\end{equation}

We define $\hlambda_t$ in terms of the above equation, but with respect to $\hmx_{\text{MAP}}$ and $\htheta$ inside the function $q_n$.
We rescale the RR intervals using
\begin{equation}
    \tau_n = \int_{u_{n}}^{u_{n+1}} \lambda_t dt
\end{equation}
and we define $\htau_n$ analogously with respect to $\hlambda_t$.

According to the time-rescaling theorem, $(\tau_n, n \geq 0)$ are i.i.d. unit-rate, exponential random variables. Therefore, we compute $\htau_n$ with respect to $(\htau_n, n \geq 0)$ and evaluate goodness of fit by computing $\hZ_n = F_{\text{exp}}(\htau_n)$, where $F_{\text{exp}}(s) = 1 - e^{-s}$ is the cumulative distribution function of a unit-rate, exponential random variable. For good model fit, $(\htau_n, n \geq 0)$ should be approximately i.i.d. unit-rate exponentials, implying that $(\hZ_n, n \geq 0)$ should be approximately i.i.d. uniform $[0,1]$ random variables.
We can evaluate this with Kolmogorov-Smirnov (KS) plots and 95\% confidence intervals.

\subsection{Converting RR intervals to heartrate}
Heartrate is often a more relevant quantity to study rather than RR intervals themselves. Our distribution of RR intervals enables closed-form calculation of heartrate, which follows an inverse gamma random variable with mean:
\begin{equation}
    \mu_{H\!R}[n] = \frac{\beta^{\dagger}_n}{\hat{\alpha}-1} = \frac{c\hat{\alpha}e^{-\vh_n^T \vx_n}}{\hat{\alpha}-1},
\end{equation}
where $c = $ \SI{60}{\second/\minute} for unit conversion.

\subsection{Time-resolve directed information impacting RR intervals}
Having a statistical model allows us to apply information-theoretic methods to quantify measures such as directed information \cite{massey1990causality}. This would allow us to capture how information flow from another source, such as another organ or an external stimulus, may impact heartbeat timings. We take advantage of the time-resolved nature of our SSM to compute dynamic information measures such as the sample path causal measure (SPCM) \cite{schamberg2019measuring}. Compared to traditional measures such as transfer entropy which take an expectation over the full joint distribution, the SPCM takes advantage of the realized sample path to quantify dynamic patterns that may be otherwise averaged out. This metric is defined as:
\begin{equation}
    \SPCM{\aux{Y}}{Y}{n} = D(P_{Y_n|y^{n-1}, \aux{y}^{n-1}} || P_{Y_n|y^{n-1}})
    \label{eqn:spcm}
\end{equation}
where $D(\cdot||\cdot) \geq 0$ is the Kullback-Leibler (KL) divergence \cite{kullback1951information}. In the case that $\aux{Y}$ provides no information about $Y$ at sample index $n$, $\SPCM{\aux{Y}}{Y}{n} \approx 0$. However, if $\aux{Y}$ provides significant information about $Y$ at sample $n$, then $\SPCM{\aux{Y}}{Y}{n} >> 0$. 

Specifically, in assessing if another stimulus impacts the duration of RR intervals, we fit two models: 1) with only the autoregressive terms and 2) with the autoregressive terms (with the same model order $d$) and additionally other regressors that represent the stimulus. From these, the KL divergence can be computed in closed form, given by:
\begin{equation*}
    \SPCM{\aux{Y}}{Y}{n} = \aux{\alpha}\ln\frac{{\beta_n}}{\aux{\beta}_n} - \ln\frac{\Gamma(\alpha)}{\Gamma(\aux{\alpha})} + (\alpha-\aux{\alpha})\psi(\alpha)-(\beta_n-\aux{\beta}_n)\frac{\alpha}{\aux{\alpha}}
\end{equation*}
where $\Gamma$ is the gamma function and $\psi$ is the digamma function. $Y_n$ and $\aux{Y}_n$ follow gamma distributions parameterized by $\alpha, \beta_n$ and $\aux{\alpha}, \aux{\beta}_n$, respectively.

\subsection{Statistical tests for respiratory sinus arrhythmia}
\label{sec:rsatests}
In this paper, we use the SPCM to assess if respiration plays a causal role in modulating RR intervals, a well-studied phenomena known as respiratory sinus arrhythmia (RSA) \cite{yasuma2004respiratory}. This effect describes the shortening of RR intervals shorten during inhalation and the lengthening during exhalation, which help increase the efficiency of ventilation. Since efferent cardiac vagal activity plays a large role in this process by transmitting information from respiratory and circulatory centers in the brainstem to modulate the firing rate of cardiac pacemaker cells, RSA has been used to assess vagal function.

To quantify this interaction, we let $Y$ be the RR intervals. For $\aux{Y}$, we let it be the sine and cosine of the respiratory phase at each QRS peak of the ECG. This allows us to capture its circular nature and allow the phase to wrap around from $0$ to $360\degree$. A diagram depicting this procedure is shown in \cref{fig:fantasiaFigure}A. The respiration cycle is not sinusoidal and includes pauses that leads to asymmetries in breathing rate \cite{diez2025unmasking}. Because of this, phases extracted with a traditional Hilbert transform method may not consistently line up with physiological markers such as the start of inhale and exhale. Instead, we use the method proposed in \cite{hathaway2025novel}. We first identify the respiration peaks and troughs using NeuroKit2 \cite{Makowski2021neurokit}. Then, respiration trough, or the beginning of inhalation, is set as $0\degree$ while respiration peak, or the beginning of exhalation, is set as $180\degree$. Between these points, the phase is linearly interpolated. 

From there, the model fitting procedure is applied to just the RR intervals. Through this, the ideal model order is selected. Using this same model order, the model is refit, this time also including the auxiliary regressors $\aux{Y}$. From there, the SPCM $\SPCM{\aux{Y}}{Y}{t}$ can be computed based on \cref{eqn:spcm}.

We assess if there is statistically significant information flow from respiration phase to RR intervals using permutation testing. To preserve the continuity in the respiration phase but to decouple it from the RR intervals, we generate a null distribution by randomly selecting $1000$ shifts $S_m$, where $m = 1,\ldots,1000$, that are each at least $\pm60$ indices. This guarantees that each shift is at least 3 respiration cycles away (as most human respiratory cycles are at most \SI{20}{\second} long) to disrupt any respiration-cardiac coupling. For each shift we slide the respiration phases by $S_m$ RR intervals, refit the model with the new respiration phases, and recompute the SPCM. Based on this procedure, we compute a p-value by counting how many of these simulations resulted in a time-averaged SPCM that is greater than the observed (unshifted) respiration.

In addition, because our model can capture temporal dynamics, we can use a linear mixed-effects model (LMEM) to characterize population-level variation in how the SPCM changes across the respiration cycle and assess whether participant demographics, such as age, modulate this relationship. The LMEM is modeled as:
\begin{equation}
     y_{ij} = \beta_0 + \ubeta_\phi^T \begin{bmatrix}
        \sin \phi_{ij} \\
        \cos \phi_{ij}
    \end{bmatrix}
    + \beta_{A} A_j \nonumber + A_j\ubeta_{A,\phi}^T \begin{bmatrix}
        \sin \phi_{ij} \\
        \cos \phi_{ij}
    \end{bmatrix}
    +b_j+\epsilon_{ij}
    \label{eqn:LMEM}
\end{equation}
where $i$ indexes the RR intervals (effectively the time samples) while $j$ indexes the different subjects. $A_j$ is the age in years, and $\phi_{ij}$ captures the phase time series for each subject. The coefficients $\beta_0$, $\ubeta_\phi$, $\beta_A$, and $\ubeta_{A,\phi}$ capture the fixed effects associated with the intercept, respiration phase, age, and the interaction between the two, respectively. $b_j$ captures subject-level random effects, while $\epsilon_{ij}$ captures the residual error for each sample.

\section{Results}
\label{sec:results}
\subsection{Synthetic heartbeats under constant conditions}
\begin{figure}[t]
    \centering
    \includegraphics[width=\columnwidth]{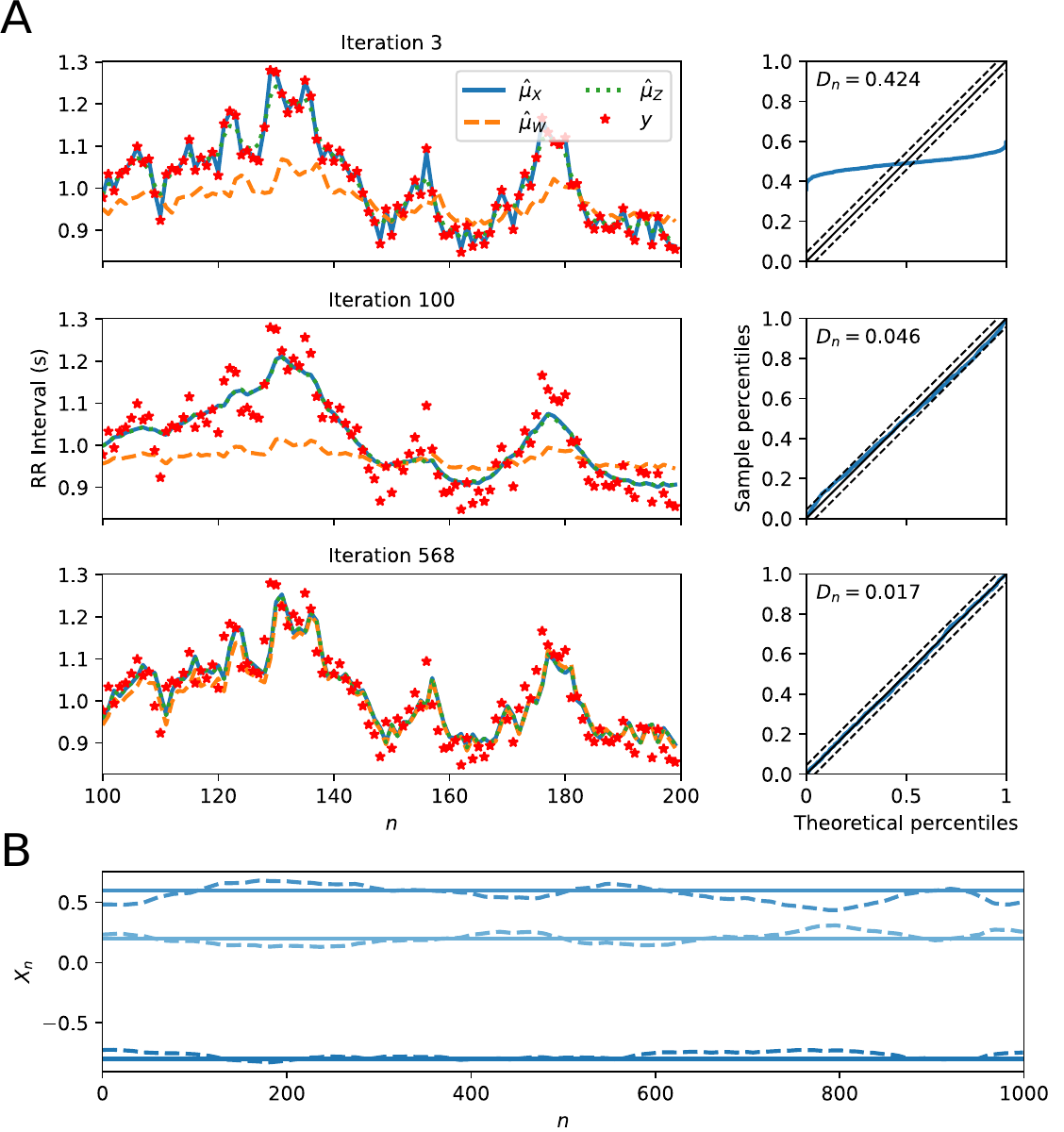}
    % \includesvg[width=\columnwidth]{figures/synthConst_combined.svg}
    \caption{Model fit from synthetic data with constant weights. Top plot shows the estimates of the mean RR interval from iterations 3, 100, and 568 which is the last iteration. The red stars show the observed RR intervals while the solid blue line shows the mean estimate from $\mX$, the dashed orange line shows the mean estimate from $\matr{Y}$, and the dotted green line shows the mean estimate from $\matr{Z}$. To the right of each plot is also the KS plot showing how model fit changes as ADMM progresses. Below are the final estimates of the weights in dashed lines with the actual weights used to generate the data in solid lines. The darkest blue corresponds to $\vx_0$ and the lightest blue corresponds to $\vx_2$.}
    \label{fig:synthConstFigure}
\end{figure}

We begin by analyzing the behavior of the latent weight estimation procedure when provided synthetically generated heartbeats. We simulate $1000$ heartbeats using a model order of $d = 2$ using constant weights $\vx = [-0.8, 0.6 ,0.2]$. This provides a mean RR interval of \SI{1}{\s} and HRV of \SI{50}{\ms}. We provide the model parameters and apply ADMM to estimate $\mX$.

\Cref{fig:synthConstFigure} shows the simulation results. The top figure shows the mean RR interval estimated from the ADMM variables $\mX$, $\matr{Y}$, and $\matr{Z}$ relative to the observed RR intervals. As the model fitting procedure progresses, we see that the procedure ends when convergence is reached between these three sets of variables. In addition, the KS plots generated by the estimates from $\mX$ are also shown on the right. Model fit improves (demonstrated by the shrinking KS distance) as the ADMM continues. The bottom plot shows the final estimates of the weights $\Hat{\mX}$ compared to the true weights that were used to generate the data $\mX$. Using this procedure, there is close agreement between the estimated and the ground truth values in both the magnitude of the weights and their relative ordering: $\vx_0 < \vx_2 < \vx_1$.

\subsection{Transient heartbeat timing changes in during tilt table tests}
\begin{figure*}[t]
    \centering
    \includegraphics[width=\textwidth]{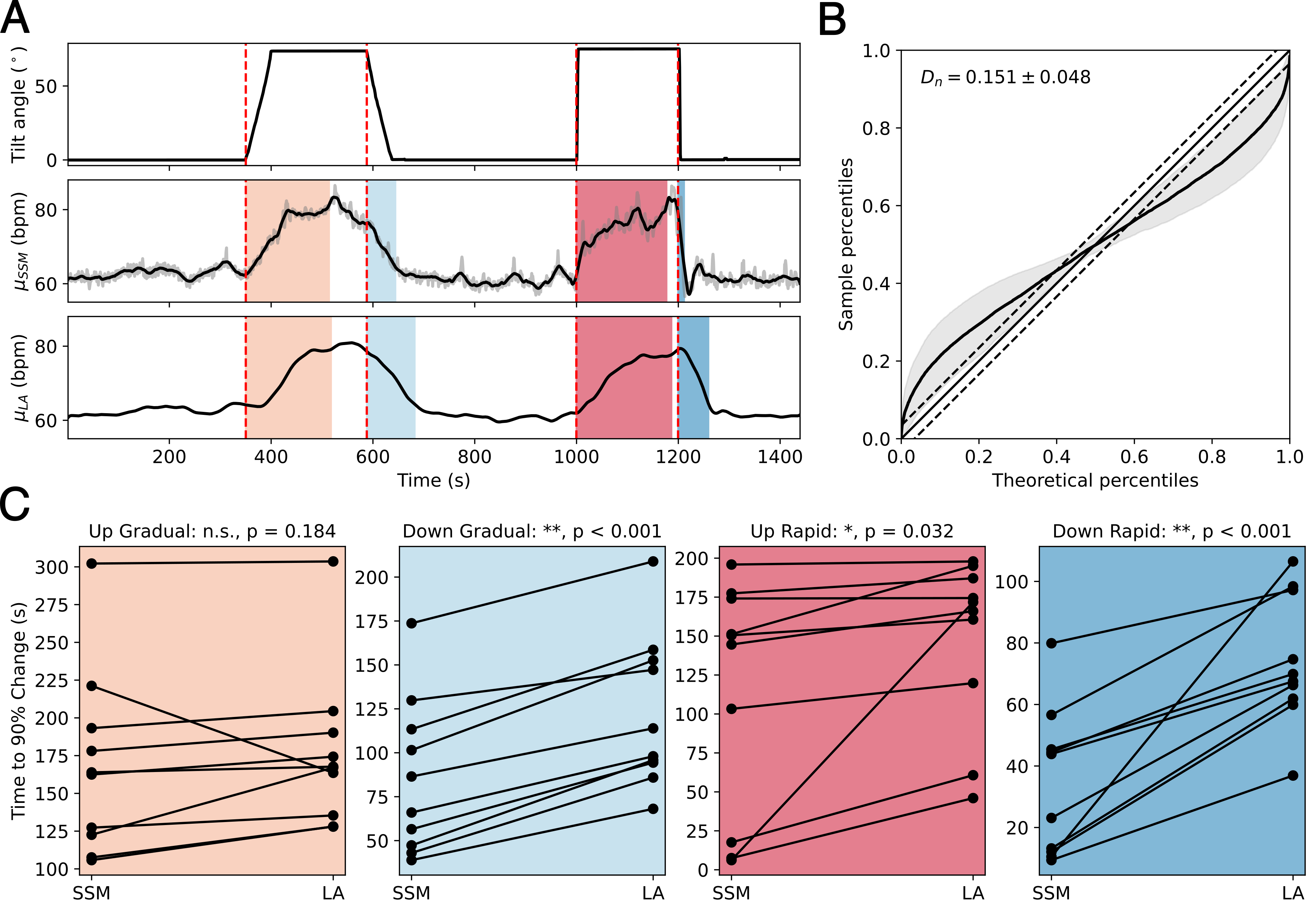}
    \caption{A) shows a sample heart rate trajectory for subject 12726 from the tilt table dataset. From top to bottom are the tilt angle, the estimated mean heartrate from the SSM, and the estimated mean heart rate from local averaging (LA). B) shows the average KS plot from all $10$ subjects. C) compares the time to $90\%$ change that can be resolved using the smoothed SSM vs. the LA trajectories. Shown p-values are after correcting for multiple comparisons using Benjamini-Hochberg.}
    \label{fig:tiltTableFigure}
\end{figure*}

The tilt table test is a commonly used clinical test to probe the ANS through postural changes. We test our procedure against publicly available tilt table data from $10$ healthy subjects on PhysioNet \cite{heldt2003circulatory, goldberger2000physiobank}. The gender distribution in this dataset is evenly split with mean age reported as $28.7\pm 1.2$ years; additional subject information can be found in \cite{heldt2003circulatory}. We consider a subsection of each subject's recordings to specifically compare their heartbeat dynamics during a slow tilt from horizontal to $75\degree$ and back versus a rapid tilt along the same trajectory. We demonstrate adequate model fit in \cref{fig:tiltTableFigure}B with the average KS distance being $0.151 \pm 0.048$ across the $10$ subjects.

In \cref{fig:tiltTableFigure}A, we show a sample HR trajectory of a subject from the dataset. In it, we see that the mean HR computed from local averaging smooths out sharp transient changes caused by the stimulus. This is especially apparent in the rapid tilt down where our model is able to capture a sharp drop in HR that is shown as more gradual in the local averaging method.

To quantify this difference, in \cref{fig:tiltTableFigure}C, we compare how much time the two methods report for the subject to reach a set amount of cardiac change for each of the four different tilts. To do this, we apply a Savitzky-Golay filter of window length $500$ and polynomial order $5$ to the estimates of both methods to capture underlying trends \cite{savitzky1964smoothing}. We then use the first \SI{5}{\min} before the first tilt as the subject's baseline heart rate, and for each tilt, find the maximum HR they achieve after the tilt up. The time that it takes from the filtered estimates to reach $90\%$ of the maximum change (either increase or decrease) is recorded. For each type of tilt, we then perform a paired, one-sided t-test, correct for multiple comparisons using the Benjamini-Hochberg (BH) procedure \cite{benjamini1995controlling}, and show that for the rapid tilt up and both tilt downs our model captures significantly faster HR change times in the subjects than in traditional local averaging methods.

\subsection{Inter-organ coupling in the respiratory-cardiac setting}
\begin{figure*}[t]
    \centering
    \includegraphics[width=\textwidth]{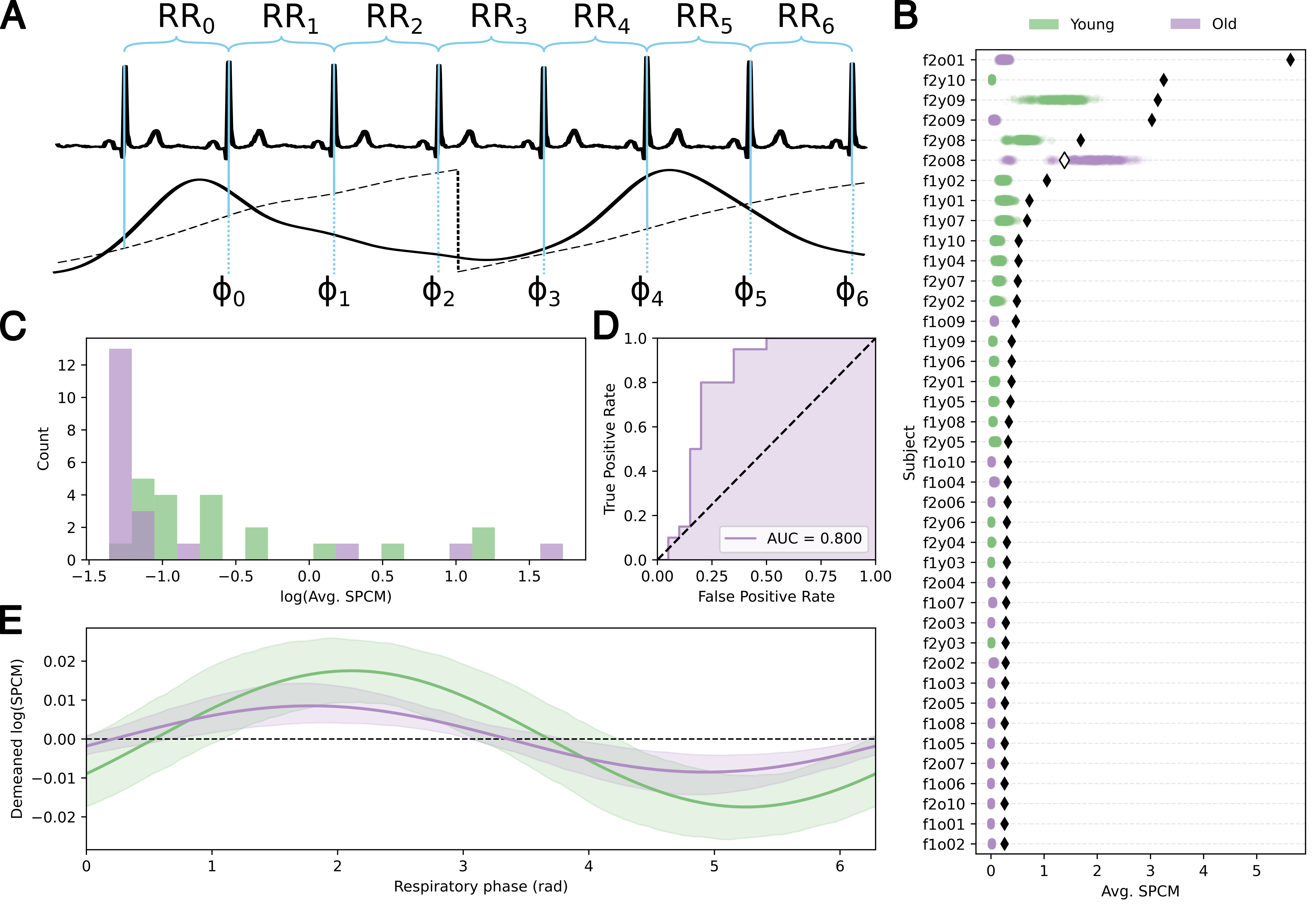}
    \caption{A) To compute the SPCM, we extract the RR intervals and the respiration phase that corresponds to when the heartbeats occur. B) Permutation testing was performed to assess if there was significant information flow. For each subject, samples from the null distributions are shown by the colored clusters while the observed sample is shown by the diamond. After correcting for multiple comparisons, subjects with significant average SPCM are indicated by a black diamond. C) Histogram showing the log of the time-averaged SPCM for the young vs. old group. D) ROC from a threshold classifier for old vs. young based on time-averaged SPCM. AUC of $0.800$. E) Demeaned log(SPCM) predicted by the LMEM for the median ages of the young ($27$ years) vs. the old ($76$ years) groups. $95\%$ confidence intervals were computed via bootstrapping.}
    \label{fig:fantasiaFigure}
\end{figure*}

One power of having a statistical heartbeat dynamics model is we can use information theoretic measures to characterize the relation between the heart and other organs, such as the lungs in RSA. It has been shown that the phase of respiration can shorten (during inspiration) or lengthen (during expiration) interbeat cardiac intervals \cite{yasuma2004respiratory}. In addition, this effect has been found to be attenuated as individuals age due to a decline in the neural component of the baroreflex that limits how changes in pressure modulate vagal outflow \cite{kaushal2002inter}.

We study this effect using simultaneous ECG and respiration recordings from $40$ healthy individuals from the publicly available Fantasia dataset \cite{iyengar1996age, goldberger2000physiobank}. Of these participants, $20$ are part of the young group aged $21$-$34$ years old while the other half are in the old group aged $68$-$81$ years old. All participants were awake, supine, and at rest for the two hour recording session. Further details about the participants can be found in \cite{iyengar1996age}. 

\cref{fig:fantasiaFigure}A shows the conceptual diagram of how we 
extract RR intervals and obtain respiratory phase from which we derive their inter-relation using the LMEM described in \eqref{eqn:LMEM}. \cref{fig:fantasiaFigure}B shows the observed average SPCM for each subject compared to the null distributions generated from permutation tests. After correcting for multiple comparisons across subjects using the BH procedure, we find that $39$ out of the $40$ subjects have significant directed information from respiration phase to RR interval.

The magnitude of the time-averaged SPCM can be used to differentiate between our young and old groups. In \cref{fig:fantasiaFigure}C, the histograms show that the young group has generally higher time-averaged SPCM than the old group. In \cref{fig:fantasiaFigure}D, we show that a classifier that simply thresholds the time-averaged sample path causal results in an area under the receiver operating characteristic curve (AUROC) of $0.800$ in classifying whether a subject falls in the young or old age range.

In this model, while the effect of age alone on the log-SPCM did not reach significance, the effects of respiration phase as well as the interactions between respiration and age did. The coefficients and associated statistics of the fitted LMEM are presented in \cref{tab:lmemParams}.

In \cref{fig:fantasiaFigure}E, we use the LMEM and predict the SPCM across the respiratory cycle for the median ages of the young ($27$ years) and old ($76$ years) groups. We demean the log-transformed SPCM and plot the predicted deviations from the average as well as the $95\%$ confidence intervals which are generated from $500$ bootstrapped samples. We show that after we remove the baseline amplitude effects of age, there are also phase-dependent effects as well. We see that in the predicted young subject, on average, the log-SPCM oscillates with an amplitude of $0.0175$ while in the predicted old subject, it oscillates with an amplitude of $0.00852$, resulting in a $51.4\%$ decrease in modulation. In addition, there was on average a $18\degree$ phase shift, with the greatest log-SPCM being concentrated at $120\degree$ in the predicted young subject and at $102\degree$ in the predicted old subject.

\begin{table*}[t]
    \centering
    \begin{tabularx}{\textwidth}{
        |>{\centering\arraybackslash}X|
        >{\centering\arraybackslash}X|
        >{\centering\arraybackslash}X|
        >{\centering\arraybackslash}X|
        >{\centering\arraybackslash}X|
        >{\centering\arraybackslash}X|
    }
        \hline
        Effect & $\beta$ & SE & p-value & $95\%$ CI Low & $95\%$ CI High \\
        \hline
        Intercept
            & $-0.424$
            & $0.271$
            & $0.118$
            & $-0.956$ & $0.108$ \\
        Age
            & $\num{-7.98e-3}$
            & $\num{4.85e-3}$
            & $0.0997$
            & $-17.5\num{e-3}$ & $1.52\num{e-3}$ \\
        $\sin\phi$
            & $0.0187$
            & $\num{2.05e-3}$
            & $<\num{1e-6}$
            & $0.0147$ & $0.0227$ \\
        $\cos\phi$
            & $-0.0129$
            & $\num{2.05e-3}$
            & $<\num{1e-6}$
            & $-16.9\num{e-3}$ & $-8.88\num{e-3}$ \\
        $\sin\phi \times \mathrm{Age}$
            & $\num{-1.37e-4}$
            & $\num{3.7e-5}$
            & $\num{2.44e-4}$
            & $-21.0\num{e-5}$ & $-6.4\num{e-5}$ \\
        $\cos\phi \times \mathrm{Age}$
            & $\num{1.46e-4}$
            & $\num{3.7e-5}$
            & $\num{8.6e-5}$
            & $7.3\num{e-5}$ & $21.9\num{e-5}$ \\
        \hline
    \end{tabularx}

    \caption{Coefficients and statistics from a linear-mixed effects model testing the effects of age and respiration phase on SPCM.}
    \label{tab:lmemParams}
\end{table*}

\section{Discussion}
\label{sec:discussion}
In this work, we propose a new convex, state-space model of heartbeat dynamics along with a procedure for how to solve for the model parameters and perform model selection. Compared to prior work, this method provides an efficient way to capture time-varying changes in the statistics of heartbeat timings using a point-process model. We demonstrate in a series of synthetic heartbeats that we can use this method to recover underlying dynamics and show in two clinical datasets that it provides physiologically meaningful insights. Together, these results show that the proposed model can provide a means to study autonomic regulation of the cardiovascular system in varying contexts.

By using gamma emissions with a log-link function, this model is convex. This helps provide theoretical guarantees on estimating the latent weights after the model parameters are specified. By adding a Gauss-Markov process prior, we are able to model changes in the underlying physiology that may occur in a variety of settings. Since the computational complexity of solving this convex optimization problem increases with the number of latent  weights to be estimated, we use ADMM paired with the Lambert function to develop an iterative solver with closed-form updates that is theoretically guaranteed to converge to the optimal estimate and efficiently solves for these weights when the other parameters are provided.

Using the simulated heartbeat series, we are able to control the underlying weights used to generate the RR intervals. Using this data, we are able to demonstrate, across iterations, the estimated weights from ADMM will improve in model fit by balancing the influence from the observations as well as the GMP priors. This is seen by the reduced KS distance. In addition, we demonstrate that the final weight estimates after reaching convergence closely match the true weights for the data.

The tilt table dataset provides evidence that compared to traditional windowing methods like local averaging that are currently used in commercial and clinical applications, such dynamic models may be more effective at capturing sharp transients in heart rate data. Specifically in the two tilt down cases (gradual and rapid) and the rapid tilt up, there was a statistically significant decrease in the time it took to achieve a $90\%$ change in heart rate measured by the two methods. In addition, we can qualitatively observe this in the example heart rate trajectories from a sample tilt table subject where we see the sharp drop in heart rate after the rapid tilt down is spread out over time in local averaging. Accurately capturing these differences are important in clinical contexts. Prior research has demonstrated that heart rate recovery after physical stress may be an important indicator of mortality \cite{cole1999heart}, while \cite{mezzacappa2001vagal} has demonstrated that similar is true after exposure to psychological stress. 

We also demonstrate the use case of applying these models to multimodal data to capture cross-organ interactions. Specifically, using the respiration and ECG recordings from the Fantasia dataset \cite{iyengar1996age}, we demonstrate that the SPCM can identify significant respiratory-cardiac coupling. In addition, we show that the model can identify age-related differences in RSA, with on average a greater influence of respiratory phase on RR intervals in the young cohort compared to the old cohort. This is in line with prior work which has shown that the strength of RSA decreases with age \cite{kaushal2002inter}. Whereas the age-related impact on the baseline SPCM did not reach statistical significance in the LMEM, the 
coefficients association with respiration phase along with the coefficients associated with the interaction between age and respiration phase were statistically significant. This suggests that we can identify a time-varying information flow between organs which change as individuals age. Compared to recent approaches that have identified organ-specific, blood plasma biomarkers of aging \cite{oh2023organ}, our work provides potential complementary avenues to characterize physiological aging by studying dynamics in multi-organ systems.

The approach and results in this manuscript highlight one benefit of such a statistical framing. Defining a probabilistic model allows us to implement information-theoretic quantities to assess changes in these values. Specifically, because we used a gamma distribution which is part of the exponential family, this was easy to compute and had closed form expressions based on its parameters. This applies generally to other members of the exponential family as well and provides opportunities for other, similar models to utilize these techniques to characterize multi-organ relations.

There are a few limitations of our method. While the Expectation-Maximization (EM) algorithm \cite{dempster1977maximum} is one procedure that has been described to estimate parameters such as the latent covariance matrices and the shape parameter, which can heavily influence model fit, we found that its usage was unable to robustly estimate the expectation step and ensure convergence. In addition, model selection for dynamic state-space models with non-Gaussian emissions is an area for improvement as well. We find trends in the PACF based on prior work in GARMA models \cite{benjamin2003generalized}, but future research into a broader class of models is needed to find theoretically optimal methods for our use case.

Beyond model fitting, there are some limitations of computing the information theoretic quantities we have proposed. In order to assess significance, because the underlying null distribution is unknown, we utilize permutation testing to generate samples to do so. This is quite computationally expensive to generate all the samples. In addition, when correcting for multiple comparisons, it is important to note that for the number of samples we generated per null distribution, we are unable to get precise p-values that are below $0.001$. 

Future work can extend this model to capture a richer class of dynamics in a broader range of cardiovascular settings. For example, expanding the model to include nonlinear changes in the underlying latent process would capture more flexible dynamics. We have also shown a way to quantify RSA using our model and determine how it changes with age. This could in the future be extended to a biomarker of autonomic aging that can be noninvasively measured through any modalities that can simultaneously record respiration and cardiac activity.

\section{Conclusion} % limit to 300 words
This work proposes a time-varying model of heartbeat dynamics. It treats heartbeat timings as a point process that is influenced by previous heartbeats and other stimuli. By combining a GMP with gamma emissions, this model leverages convexity for computational efficiency to solve for the optimal high-dimensional latent weights. This is important for applications that require analyzing dynamics over longer time periods, such as out-of-clinic ambulatory settings. We demonstrate that the beat-by-beat time resolution of our model can captures sharp transient changes in heart rate dynamics that may be otherwise smoothed over and missed in traditional windowing methods. We demonstrate that this is especially relevant in measuring heart rate recovery times after postural stress, which can be an important indicator of autonomic health. The structure of the model also allows us to look at how other stimuli, such as respiratory phase, can modulate heartbeat timings by leveraging information-theoretic measures. We show how we can quantify the strength of RSA with this model which can potentially serve as a biomarker of autonomic health and aging. Together, this work proposes a heartbeat dynamics model that is both practical to compute because of its computational efficiency and useful in its ability to capture dynamics that may otherwise be missed by windowing methods or static models.

\section*{Acknowledgments}
S.L. would like to acknowledge support of this work by the Robert Bosch Stanford Graduate Fellowship and the Northern California Chapter of the ARCS Foundation. A.S.P would like to acknowledge support of this work by Stanford Data Science, the Siebel Foundation, and NIH T32CA272386. T.P.C. would like to acknowledge support of this work by the Wu Tsai Human Performance Alliance at Stanford University, the Joe and Clara Tsai Foundation, the Wu Tsai Neurosciences Institute, the Yu Faculty Scholar funds from the Stanford School of Engineering, the Stanford Bio-X Interdisciplinary Initiatives Seed Grants Program, and NIH 1U01DK140939. 

%Bibliography
\bibliographystyle{unsrt}  
\bibliography{refs}

\end{document}